\documentclass[preprint,prd,aps,nofootinbib]{revtex4}
\usepackage{amsmath}
\usepackage{psfig}
\newcommand{\be}{\begin{equation}}
\newcommand{\ee}{\end{equation}}
\newcommand{\ba}{\begin{eqnarray}}
\newcommand{\ea}{\end{eqnarray}}
\newcommand{\bc}{\begin{center}}
\newcommand{\ec}{\end{center}}

\begin{document}
\title{Arnowitt--Deser--Misner gravity with variable $G$ and $\Lambda$ 
and fixed point cosmologies from the renormalization group}
\author{Alfio Bonanno,$^{1,2}$
\thanks{Electronic address: abo@ct.astro.it}
Giampiero Esposito$^{3,4}$
\thanks{Electronic address: giampiero.esposito@na.infn.it}
and Claudio Rubano$^{4,3}$
\thanks{Electronic address: claudio.rubano@na.infn.it}}
\address{${ }^{1}$Osservatorio Astrofisico, Via S. Sofia 78,
95123 Catania, Italy\\
${ }^{2}$Istituto Nazionale di Fisica Nucleare, Sezione di Catania,\\
Corso Italia 57, 95129 Catania, Italy\\
${ }^{3}$Istituto Nazionale di 
Fisica Nucleare, Sezione di Napoli,\\
Complesso Universitario di Monte S. Angelo, Via Cintia,
Edificio N', 80126 Napoli, Italy\\
${ }^{4}$Dipartimento di Scienze Fisiche,
Complesso Universitario di Monte S. Angelo,\\
Via Cintia, Edificio N', 80126 Napoli, Italy}

\vspace{2cm}
\begin{abstract}
Models of gravity with variable $G$ and $\Lambda$ have acquired
greater relevance after the recent evidence in favour of
the Einstein theory being non-perturbatively 
renormalizable in the Weinberg sense. The present paper applies
the Arnowitt--Deser--Misner (ADM) formalism to 
such a class of gravitational models. A modified action functional
is then built which reduces to the Einstein--Hilbert action when
$G$ is constant, and leads to a power-law growth of the scale factor
for pure gravity and for a massless $\phi^4$ theory in a Universe
with Robertson--Walker symmetry, 
in agreement with the recently developed 
fixed-point cosmology. Interestingly, the renormalization-group flow
at the fixed point is found to be compatible with a Lagrangian
description of the running quantities $G$ and $\Lambda$.
PACS: 04.20.Fy, 04.60.-m, 11.10.Hi, 98.80.Cq
\end{abstract}
\pacs{04.20.Fy, 04.60.-m, 11.10.Hi, 98.80.Cq}
\maketitle
\section{Introduction}
Recent studies support the idea that the 
Newton constant $G$ and the cosmological 
constant $\Lambda$ are actually spacetime functions 
by virtue of quantum fluctuations 
of the background metric \cite{[1],[2],[3],[4]}. Their behaviour is ruled 
by the renormalization group (hereafter RG) 
equations for a Wilson-type Einstein-Hilbert
action where $\sqrt{g}R$ and $\sqrt{g}$ become 
relevant operators in the neighbourhood of 
a non-perturbative ultraviolet fixed point 
in four dimensions \cite{[5]}. The theory is thus 
{\it asymptotically safe} in the Weinberg sense \cite{[6]} because the
continuum limit is recovered at this new ultraviolet fixed point. 
In other words, the theory is non-perturbatively renormalizable
\cite{[7],[8],[9],[10]}.

Within this framework, the basic ingredient to promote 
$G$ and $\Lambda$ to the role of spacetime functions 
is the {\it renormalization group improvement},
a standard device in particle physics in order to add, 
for instance, the dominant 
quantum corrections to the Born approximation of a scattering cross section. 
The basic idea of this approach is similar
to the renormalization  group based
derivation of the Uehling correction to the Coulomb potential in
massless QED \cite{[11]}. 
The ``RG improved'' Einstein equations can thus be obtained 
by replacing $G \to G(k)$, $\Lambda \to \Lambda(k)$, 
where $k$ is the running mass scale 
which should be identified with 
the inverse of cosmological time in a homogeneous 
and isotropic Universe, $k \propto 1/t$
as discussed in Refs. \cite{[3],[4]}, or with inverse of 
the proper distance $k \propto 1/d(P)$
of a freely falling observer in a Schwarzschild background \cite{[1],[2]}.
 
In this way the RG running gives rise to a dynamically evolving, 
spacetime dependent $G$ and $\Lambda$. 
The improvement of Einstein's equations can then 
be based upon any RG trajectory $k \mapsto (G(k), \Lambda(k))$ 
obtained as an (approximate) solution to the 
exact RG equation of (quantum) Einstein-Gravity.
Within this framework it has been shown that 
the {\it renormalization group derived cosmologies} 
provide a solution to the horizon and 
flatness problem of standard cosmology without any inflationary mechanism. 
They represent also a promising model of dark energy 
in the late Universe \cite{[12]}.
A similar approach has also been discussed in Ref. \cite{[13]}, where the 
RG equation arises from the matter field fluctuations. 

In comparison to earlier work
\cite{[14],[15],[16],[17],[18]} on cosmologies 
with a time dependent $G$, $\Lambda$
and possibly fine structure constant $\alpha$ the new feature of these
RG derived cosmologies is that
{\it the time dependence of $G$ and $\Lambda$ is a 
secondary effect which results from a
more fundamental scale dependence}. In a typical 
Brans-Dicke type theory the dynamics
of the Brans-Dicke field $\omega = 1/G$ is governed 
by a standard local Lagrangian with
a kinetic term $\propto (\partial_\mu \omega)^2$. 
In this approach there is no simple
Lagrangian description of the $G$-dynamics {\it a priori}. 
It rather arises from an RG
equation for $G(k)$ and a cutoff identification 
$k=k(x^\mu)$. From the point of view of the
gravitational field equations, $G(x^\mu)$ has the 
status of an {\it external} scalar field whose evolution is engendered by 
the RG equations. 
It is nevertheless interesting to notice that 
a dynamically evolving cosmological 
constant and {\it asymptotically free} gravitational interaction also appear
in very general scalar-tensor cosmologies \cite{[19],[20]}.  

In general, the RG equations do not admit
a gradient flow representation and it is not clear 
how to embed the RG behaviour
into a Lagrangian formalism. This problem has been widely discussed 
in Ref. \cite{[21]} where a 
consistent RG improvement of the Einstein-Hilbert action has been 
proposed at the level of the four-dimensional Lagrangian.

Instead, {\it the relevant question we would like to study in this paper 
is how to achieve a modification of the standard ADM Lagrangian
where $G$ and $\Lambda$ 
are dynamical variables according to a prescribed renormalized trajectory}
(see Ref. \cite{[22]} for a first attempt in this direction).
The simplest non-trivial renormalization group trajectory is represented 
by the scaling law near a fixed point for which 
\be\label{fixx}
G(x^\mu) \; \Lambda(x^\mu) = {\rm const},
\ee
and we  shall  use the simple relation (\ref{fixx}) 
to constrain the possible dynamics.

Indeed, a theory with an independent dynamical $G$ is known to be
equivalent to metric-scalar gravity already at classical level. In
particular, it can be reduced to canonical form with the standard 
expression of the kinetic term for the scalar by a conformal rescaling
of the metric (see a discussion of this point in Ref. \cite{[23]}). At this 
stage, a theory where Lambda is an independent dynamical variable meets
serious problems, in agreement with what we find below.

Following the fixed-point relation as implemented 
in Eq. (\ref{fixx}), in this paper we
thus discuss a dynamics according to which
$\Lambda$ depends on $G$ which is, in turn,
a function of position and time. If $\Lambda$ and $G$ were
instead taken to be independent functions of position and time,
the primary constraint of vanishing conjugate
momentum to $\Lambda$ would lead to a secondary constraint
which is very pathological, and details will be given later in 
Sec. II to avoid logical jumps.
 
In Sec. II we introduce and motivate a modified action 
functional for theories of
gravitation with variable $G$ and $\Lambda$. 
Such a result is applied, in Sec. III, to pure
gravity and to gravity coupled to a massless self-interacting scalar
field in a Universe with Robertson--Walker
(hereafter RW) symmetry.
Concluding remarks and open problems are presented in Sec. IV,
while the full Hamiltonian analysis is performed in the appendix.

\section{Modified action functional}

According to the ADM treatment of 
space-time geometry, we now
assume that the space-time manifold $(M,g)$ is topologically
$\Sigma \times R$ and is foliated by a family of spacelike 
hypersurfaces $\Sigma_{t}$ all diffeomorphic to $\Sigma$. The metric
is then locally cast in the ADM form
\begin{equation}
g=-(N^{2}-N_{i}N^{i})dt \otimes dt +N_{i}(dx^{i} \otimes dt
+dt \otimes dx^{i})+h_{ij}dx^{i} \otimes dx^{j},
\label{(2)}
\end{equation}
where $N$ is the lapse function and $N^{i}$ are the components of 
the shift vector \cite{[24]}. 
To obtain the ADM form of the action, one has
to consider the induced Riemannian metric $h_{ij}dx^{i} \otimes dx^{j}$
on $\Sigma$, the extrinsic-curvature tensor 
$K_{ij}={1\over 2N}\Bigr(-h_{ij,0}+2N_{(i \mid j)}\Bigr)$ of 
$\Sigma$ (hereafter $h_{ij,0} \equiv {\partial h_{ij} \over \partial t}$,
and similarly for $N,N_{i}$ and $G$), 
and add a suitable boundary term to the Einstein--Hilbert
action, which is necessary to ensure stationarity of the full action
functional in the Hamilton variational problem \cite{[25]}. 
More precisely, the fundamental identity in Ref. \cite{[26]}
(hereafter $h \equiv {\rm det} \; h_{ij}, K \equiv K_{\; i}^{i}$,
and ${ }^{(3)}R$ is the scalar curvature of $\Sigma$)
\begin{equation}
\sqrt{-g} \; { }^{(4)}R=N \sqrt{h}(K_{ij}K^{ij}-K^{2}
+{ }^{(3)}R)-2(K \sqrt{h})_{,0}+2f_{\; ,i}^{i},
\label{(3)}
\end{equation}
where 
\begin{equation}
f^{i} \equiv \sqrt{h}\Bigr(KN^{i}-h^{ij}N_{,j}\Bigr),
\label{(4)}
\end{equation}
suggests using the Leibniz rule to express
\begin{equation}
{1\over G}(K \sqrt{h})_{,0}
={G_{,0}\over G^{2}}K \sqrt{h}+
\left({K \sqrt{h}\over G}\right)_{,0},
\label{(5)}
\end{equation}
\begin{equation}
{1\over G}{\partial f^{i}\over \partial x^{i}}
={G_{,i}\over G^{2}}f^{i}+{\partial \over \partial x^{i}}
\left({f^{i}\over G}\right),
\label{(6)}
\end{equation}
so that division by $16 \pi G$ in the integrand of the
Einstein-Hilbert action yields the Lagrangian (the $c^{4}$
factor in the numerator is set to $1$ with our choice of units)
\begin{equation}
L={1\over 16\pi}\int \left[{N \sqrt{h}\over G}(K_{ij}K^{ij}
-K^{2}+{ }^{(3)}R-2 \Lambda)
-2{G_{,0}\over G^{2}}K \sqrt{h}
+2{G_{,i}f^{i}\over G^{2}}\right]d^{3}x,
\label{(7)}
\end{equation}
after adding to the action functional the boundary term
(cf. Ref. \cite{[25]})
$$
I_{\Sigma}={1\over 8\pi}\int_{\Sigma}{K \sqrt{h}\over G}d^{3}x,
$$
and assuming that $\Sigma$ is a closed manifold (so that the total
spatial divergence of ${f^{i}\over G}$ resulting from (3) and
(6) yields vanishing contribution, having taken 
$\partial \Sigma = \emptyset$). 

The Lagrangian (7), however, suffers from a serious drawback,
because the resulting momentum conjugate to the three-metric reads as
\begin{equation}
p^{ij} \equiv {\delta L \over \delta h_{ij,0}}
=-{\sqrt{h}\over 16 \pi G}(K^{ij}-h^{ij}K)
+{\sqrt{h}\over 16 \pi G}{h^{ij}\over NG}(G_{,0}-G_{,k}N^{k}).
\label{(8)}
\end{equation}
This would yield, in turn, a Hamiltonian containing a term quadratic
in $G_{,0}$ (since $K^{ij}$ depends,
among the others, on $p^{ij}$ and on
$G_{,0}h^{ij}$), despite the fact that (7) is only linear in
$G_{,0}$ when expressed in terms of first and second fundamental
forms. There is therefore a worrying lack of equivalence between $K^{ij}$
and $p^{ij}$. 
 
We thus decide to include 
a ``bulk'' contribution in order to cancel the effect of
$G_{,0}$ and $G_{,i}$ in Eq. (7) by writing 
\begin{equation}
I \equiv {1\over 16\pi}\int_{M}{({ }^{(4)}R-2\Lambda)\over 
G({\vec x},t)}\sqrt{-g} \; d^{4}x
+{1\over 8\pi}\int_{M}{(K \sqrt{h})_{,0}\over 
G({\vec x},t)}d^{4}x
-{1\over 8\pi}\int_{M}{f_{\; ,i}^{i}\over 
G({\vec x},t)}d^{4}x,
\label{(9)}
\end{equation}
as a starting action defining a gravitational theory with variable $G$ and 
$\Lambda$, where the added terms in Eq. (9) have integrands which
are not four-dimensional total divergences.
More precisely, upon division by $16 \pi G$ in the integrand
of the Einstein--Hilbert term (first integral in Eq. (9)), 
the second and third integral in Eq. (9)
cancel the effect of $-2(K\sqrt{h})_{,0}$ and $2f_{\; ,i}^{i}$ in Eq.
(3), respectively. The resulting Lagrangian density belongs to the
general family depending only on fields and their first derivatives,
which is the standard assumption in local field theory.
It should also be noticed that if $G$ were constant, 
the second integral on the right-hand side
of (9) would reduce to the York--Gibbons--Hawking 
boundary term \cite{[25],[27]}
$$
{1\over 8\pi G}\int_{\Sigma}K \sqrt{h} \; d^{3}x.
$$
Moreover, for a constant $G$, the third integral on the right-hand
side would reduce to
$$
-{1\over 8\pi G}\int dt \int_{\partial \Sigma}f^{i}n_{i}d^{2}x,
$$
which vanishes if $\Sigma$ is the smooth boundary of $M$
(since then $\partial \Sigma=\partial \partial M=0$). 

In other words, on renormalization-group improving the gravitational 
Lagrangian in the ADM approach, one might think that  
$G$ and $\Lambda$ have the status of given external field, 
whose evolution is in principle
dictated by the RG flow equation. However, it is 
also interesting to understand whether one can 
generalize the standard ADM Lagrangian in order to consider $G$ as a 
{\it dynamical field} and investigate which dynamics 
is consistent with the RG approach. In this
spirit we eventually consider the following general ADM
Lagrangian:
\begin{equation}
L={1\over 16\pi}\int \left[{N \sqrt{h}\over G}(K_{ij}K^{ij}
-K^{2}+{ }^{(3)}R-2 \Lambda)\right]d^{3}x +L_{\rm int}+L_{\rm matter},
\label{lag}
\end{equation}
where the first term has the same functional form as the Lagrangian
of ADM general relativity (but with $G$ and $\Lambda$ promoted to
mutually related dynamical variables),
$L_{\rm int}$ is an interaction term of a kinetic type 
which, for dimensional reasons, must be of the form
(the coefficient $16 \pi$ is introduced for later convenience)
\begin{eqnarray}\label{BD}
\; &\;& L_{\rm int}=  -{\mu \over 16 \pi}
\int  {g^{\rho \sigma}G_{;\rho}G_{;\sigma} \over G^{3}}\; 
\sqrt{-g} \; d^{3}x \nonumber \\
&=& {\mu \over 16 \pi} \int {N \sqrt{h}\over G^{3}}
\left[N^{-2}(G_{,0})^{2}-2{N^{i}\over N^{2}}G_{,0}G_{,i}
-\left(h^{ij}-{N^{i}N^{j}\over N^{2}}\right)G_{,i}G_{,j}\right]d^{3}x,
\end{eqnarray}
$\mu$ being the interaction parameter, and the occurrence 
of lapse and shift
being the effect of ADM coordinates in writing the integration measure
for the action and the space-time metric.
$L_{\rm matter}$ is the ``matter''
Lagrangian that we shall consider as given by a self-interacting
scalar field. The first line of Eq. (11) stresses that we start from
an action which is invariant under four-dimensional diffeomorphisms,
although eventually re-expressed in ADM variables (second line of
Eq. (11)). It should be emphasized that there are no observational
constraints on the term $L_{\rm int}$ in Eq. (11), since we are
considering modifications of general relativity which only occur in
the very early universe at about the Planck scale. Experimental
verifications of our model, although clearly desirable, are beyond
the aims of the present work.

Note that, if
$\Lambda$ were a variable function, but independent of $G$, the
primary constraint of vanishing conjugate momentum to 
$\Lambda$ would lead to the secondary constraint
${N\sqrt{h}/ (8\pi G)}$, which would vanish on the constraint
manifold. This is very pathological, because it implies that
either the lapse vanishes or the induced three-metric on the
surfaces of constant time has vanishing determinant. Neither of these
alternatives seems acceptable in a viable space-time model.

We have also found that, if $\mu$ is instead set to zero,
one obtains the additional primary constraint
$\pi_{G} \approx 0$ (the weak-equality symbol $\approx$ is used
for equations which only hold on the constraint 
surface \cite{[28],[29]}), and the
resulting dynamical system, with its evolution and constraint
equations, is incompatible with a dynamical evolution of the scale
factor, leading only to an Einstein-Universe type of solution. 
For this reason we can conclude that the generalized Lagrangian 
described in Eq. (10) represents a minimal viable modification of the
standard gravitational Lagrangian which would lead to a dynamical $G$. 

\section{RW symmetry}

In this section we study a class of scalar field cosmologies 
within our new modified Lagrangian framework. 
On using the ADM formalism, we take a 
scalar-field Lagrangian \cite{[30]}
\begin{equation}
L_{m} \equiv \int {N \sqrt{h}\over 2}\left[N^{-2}(\phi_{,0})^{2}
-2{N^{i}\over N^{2}}\phi_{,0}\phi_{,i}
-\left(h^{ij}-{N^{i}N^{j}\over N^{2}}\right)\phi_{,i}\phi_{,j}
-2V(\phi) \right]d^{3}x,
\label{(12)}
\end{equation}
where the potential $V(\phi)$ is, for the time being, un-determined,
and $g^{00}$ is negative with our convention for the space-time
metric. 

We focus, hereafter, on cosmological models with RW symmetry.
Strictly speaking, such a name can be criticized, since we are no
longer studying general relativity, nor are we simply RG-improving
the Einstein equations. Nevertheless, we will find that spatially
homogeneous and isotropic cosmological models of the RW class can
still be achieved. In such models with lapse function $N=1$, 
the full Lagrangian, including scalar field, reads as (here 
${\cal K}=1,0,-1$ for a closed, spatially flat or open universe,
respectively)
\begin{equation}
L={a^{3}\over 16 \pi G}\left(-6{{\dot a}^{2}\over a^{2}}
+{6{\cal K}\over a^{2}}-2\Lambda \right)
+ {\mu \over 16 \pi} {a^{3}{\dot G}^{2}\over G^3}
+a^{3}\left({{\dot \phi}^{2}\over 2}-V(\phi)\right),
\label{(13)}
\end{equation}
where hereafter we revert to dots, for simplicity, to denote
derivatives with respect to $t$. Thus, the resulting second-order
Euler--Lagrange evolution equations for $a,G$ and $\phi$ are 
\begin{eqnarray}\label{eqa}
&&{{\ddot a} \over a}
+{1\over 2} {{\dot a}^{2} \over a^{2}} 
+{{\cal K} \over 2 a^{2}}-{\Lambda \over 2}
-{{\dot a} \over a}{{\dot G} \over G}+{\mu\over 4}  
{{\dot G}^2 \over G^2} +
4 \pi G \left({{\dot \phi}^{2}\over 2}-V(\phi)\right)=0, \\[2mm]
&&
\mu{\ddot G}
-{3\over 2}\mu {{\dot G}^2\over G}
+3\mu{\dot a \over a} {\dot G}  
+{G \over 2 }\left(-6{{\dot a}^{2}\over a^{2}}
+6{{\cal K} \over a^{2}}-2\Lambda 
+2G{d \Lambda \over d G}\right)=0, \label{eqg}\\[2mm]
&&{\ddot \phi}+3{{\dot a}\over a}{\dot \phi}
+{dV \over d\phi}=0. \label{eqk}
\ea
Moreover, since the Lagrangian (10) is independent of time 
derivatives of the lapse, one has the primary constraint of vanishing
conjugate momentum to $N$ (see Eq. (A1) of the appendix). 
The preservation in time of such a primary
constraint yields, for our Lagrangian generated from the assumption 
of RW symmetry, the constraint equation (cf. Eqs. (A8) and
(A10) of the appendix)
\begin{equation}
\left({{\dot a}\over a}\right)^{2}+{{\cal K} \over a^{2}}
-{\Lambda \over 3}-{\mu \over 6}  {{\dot G}^{2} \over G^2}
-{8\pi G \over 3}\left({{\dot \phi}^{2}\over 2}+V(\phi)\right)
\approx 0 .
\label{(17)}
\end{equation}
This latter equation can be rewritten in a more familiar form by introducing 
vacuum, matter, $G$ energy densities, respectively, as follows:
\be
\rho_\Lambda \equiv {\Lambda(t) \over 8\pi G(t)}, \;\;\;\;
\rho_\phi \equiv \dot\phi^2/2+V(\phi), \;\;\;\;\;\;\
\rho_{G} \equiv { \mu \over 16\pi}  {\dot G^2 \over G^3},
\ee
so that the total energy density is given by 
$\rho=\rho_\Lambda+\rho_\phi+\rho_{G} $, and Eq. (17) reads as
\be\label{14b}
\frac{{\cal K}}{a^2}=\Big( \frac{\dot{a}}{a} \Big )^2(\Omega-1),
\ee
where $\Omega \equiv \rho /\rho_{\rm cr}$, 
and the critical density is defined as usual by
\be
\rho_{\rm cr} \equiv {3 \over 8\pi G(t)} 
\left ({\dot a  \over a }\right )^{2}.
\ee
Although it would be interesting to discuss the general 
properties of the above system, our main motivation is
{\it to use RG arguments to select a particular class of possible solutions}.
In fact the RG evolution of $G$ and $\Lambda$ near a fixed point 
strongly constrains the possible solutions. More precisely, it is assumed
that there exists a fundamental scale dependence of Newton's 
parameter which is governed by an exact RG equation 
for a Wilsonian effective action whose 
precise nature need not be specified here. At a typical 
length scale $\ell$ or mass scale 
$k = \ell^{-1}$ those ``constants''
assume the values $G(k)$ and $\Lambda(k)$, respectively. 
On trying to ``RG-improve'' $G$ and $\Lambda$
the crucial step is the identification of the scale 
$\ell$ or $k$ which is relevant for 
the situation under consideration. 
In cosmology, the postulate of homogeneity and
isotropy implies that $k$ can only depend on the cosmological 
time, so that the scale dependence is
turned into a time dependence:
\be
G(t)\equiv G(k=k(t)),\;\;\; \Lambda(t) \equiv \Lambda(k=k(t)).
\ee\label{3.d}
In principle the time dependence of $k$ can be either 
explicit or implicit via the scale factor:
$k=k(t,a(t),\dot{a}(t),\ddot{a}(t),\cdots )$. 
In Refs. \cite{[3],[4]} detailed arguments are given as to why
the explicit purely temporal dependence is
\be\label{ide}
k(t) = {\xi} /{t},
\ee
$\xi$ being a positive constant. In a nutshell,
the argument is that, when the age of the Universe is $t$, 
no (quantum) fluctuation with a frequency
smaller than $1/t$ can have played any role as yet. 
Hence the integrating-out of modes
(``coarse graining'') which underlies the Wilson 
renormalization group should be stopped 
at $k\approx 1/t$. In the neighbourhood of a fixed point 
$(g_\star,\lambda_\star)$
the evolution of the dimensionful $G$ and 
$\Lambda$ is approximately given by
\be\label{36}
G(k) = \frac{g_\star}{k^2},\;\;\;\; \Lambda(k) 
= \lambda_\star \; k^{2}.
\ee
From
(\ref{36}) with (\ref{ide}) we obtain the time-dependent 
Newton parameter and cosmological term:
\be\label{9}
G(t) = g_\star \xi^{-2} \; t^{2},\;\;\;\; \Lambda(t) 
= \frac{\lambda_\star \xi^2}{t^2}.
\ee
We should mention, however, that there is another choice in the
literature, where $G$ and $\Lambda$ are related, through the
renormalization group, to the Hubble parameter $H$. Since $H$ is 
a function of the metric which is directly related to the energy
of gravitational quanta for the cosmological setting, such a
choice has been viewed to fit more naturally with the RG approach by
some authors \cite{[31]}.

The power laws (24) are valid for 
$t \rightarrow 0$ (UV case, Early Universe)
or for $t\rightarrow \infty$ (IR case, Late Universe), respectively. 
If we use these functions $G(t)$ and $\Lambda(t)$ in the dynamical system 
its solution gives us the scale factor 
$a(t)$ and the density $\rho_\phi (t)$ of the  
``RG improved scalar field cosmology''. Let us now discuss some solutions.

\subsection{Pure gravity}

Unlike models where only the Einstein equations are RG-improved,
our framework allows for a non-trivial dynamics of the scale factor
even in the absence of coupling to a matter field.
To appreciate this point, consider first the case
when no scalar field exists, so that Eq. (\ref{eqk}) 
should not be considered. 
The relation (\ref{36}) suggests looking for power-law solutions of the type
\begin{equation}\label{eq:power}
a(t)=A \;t^\alpha, 
\;\;\;\;\; G(t)=  g_\star \xi^{-2} t^{2},\;\;\;\; 
\Lambda(t)= \lambda_\star \xi^2 t^{-2},
\end{equation}
and separately consider the ${\cal K}=0$ and ${\cal K}=\pm 1$ case. 
For ${\cal K}=0$ we obtain that $A$ is an un-determined factor, while 
\be\label{nopk0}
\mu_\pm = \frac{1}{4} (3\pm\sqrt{9+12 \;\xi^2\lambda_\star}) , \;\;\;\; 
\alpha_\pm = \frac{1}{6}(3\pm\sqrt{9+12\;\xi^2\lambda_\star}),
\ee
which implies a power-law inflation for the ``$+$'' solution, 
$\alpha_+$ being larger than $1$ if $\lambda_\star>0$, 
and a possible solution of the horizon problem. 
Note that the first equality is 
a relation between coupling constants which has to be 
satisfied, while the second simply relates the
value of $\alpha$ with the product $\xi^2 \lambda_\star$. 
Since $\xi$ is not determined,
$\alpha_+$ can be made arbitrarily large. Moreover, 
both $\Omega_{G}$ and $\Omega_\Lambda$
are constant, since 
\be
\Omega_{G} = 1-\Omega_\Lambda = \frac{6}{3\pm\sqrt{9+12 \; 
\xi^2\lambda_\star}}.
\ee

If ${\cal K}=\pm 1$ we find instead $\alpha =1$ and 
\be\label{nopk1}
\mu = \frac{1}{4}(6+\xi^2\lambda_\star) , \;\;\;\;\;\;\; 
\frac{{\cal K}}{A^2} = \frac{\xi^2\lambda_\star}{2},
\ee
where, as before, the former equation relates the values of 
the coupling constants, while the latter 
is a consistency relation. In particular we see that, if 
${\cal K}=-1$, then $\lambda_\star$ must be negative.

In both cases (i.e. ${\cal K}=\pm 1$),  
$\Omega_\Lambda$ and $\Omega_{G}$ are constant with 
\be
\Omega_\Lambda = \frac{2{\cal K}}{3 A^2}, \;\;\;\;\; 
\Omega_{G} = 1+\frac{{\cal K}}{3A^2}.
\ee

\subsection{Inclusion of a scalar field}

Here we consider the contribution of a scalar field with 
a self-interacting potential of the type 
\be
V(\phi) = \frac{\zeta}{4!}\phi^{4},
\ee
with the ansatz (\ref{eq:power}) for $a,G,\Lambda$,  
and $\phi = \phi_0 t^{-\beta}$ for the scalar field. The
Klein--Gordon equation of motion (16) then yields $\beta=1$ and
\be\label{eqkk}
\phi_0= \pm \sqrt{\frac{6(3\alpha -2)}{\zeta}},
\ee
which implies $\alpha>2/3$ so as to have a real scalar field. 
We then find, if ${\cal K}=0$,
\ba\label{a1}
&&\alpha_{\pm} = \mu \pm \sqrt{\mu^2-\frac{4\mu}{3}
-\frac{2\xi^2\lambda_\star}{3} }, \\[2mm]
&&\zeta_\pm = \frac{ 
24 g_\star\pi[3\xi^2\lambda_\star-3\mu^{2}+5\mu 
\pm (1-\mu)\sqrt{3\mu (3\mu-4)-6\xi^2\lambda_\star}]} 
{\xi^2(3\xi^2\lambda_\star +2(3-2\mu)\mu)} , 
\label{con}
\ea
where (\ref{con}) is a consistency relation, 
(\ref{a1}) determines the value of $\alpha$, and a 
reality condition for $\xi^2\lambda_\star$ is given by 
$(3\xi^2\lambda_\star +2(3-2\mu)\mu)>0$. 
It is not difficult to see that there are physically interesting 
solutions with power-law inflation if
$\mu$ is large enough and positive, and $\xi$ is positive. 
A plot of the behaviour of $\alpha$ and $\zeta$ as a function of 
$\mu$ and $\xi$ is depicted 
in Figs. (\ref{fig1}) and (\ref{fig2}), respectively, where the 
fixed-point values 
$g_\star=0.31$ and $\lambda_\star = 0.35$ have been calculated 
in Ref. \cite{[32]} for a self-interacting scalar
field in Einstein--Hilbert gravity. 

\begin{figure}[!h]
\centerline{\hbox {\psfig{figure=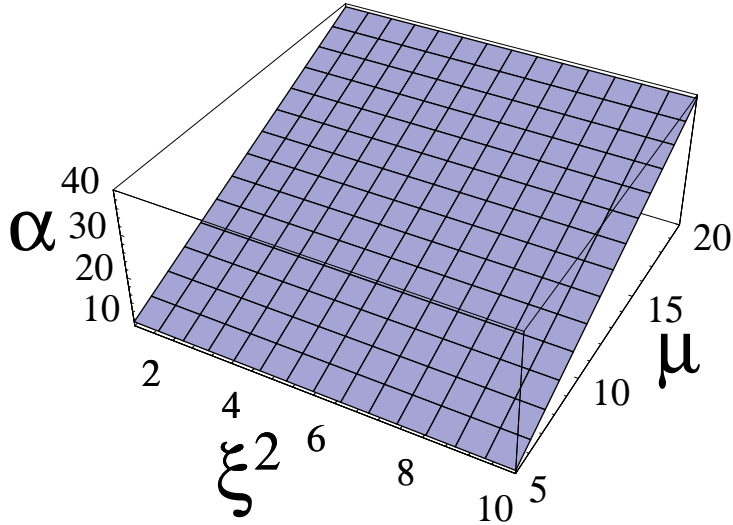,width=0.6\textwidth}}}
\caption{The exponent $\alpha$ as a function of $\mu$ and $\xi$ 
for the fixed-point values
$g_\star=0.31$ and $\lambda_\star = 0.35$ evaluated from the 
RG equation for a scalar field. 
\label{fig1}}
\end{figure}

\begin{figure}[!h]
\centerline{\hbox {\psfig{figure=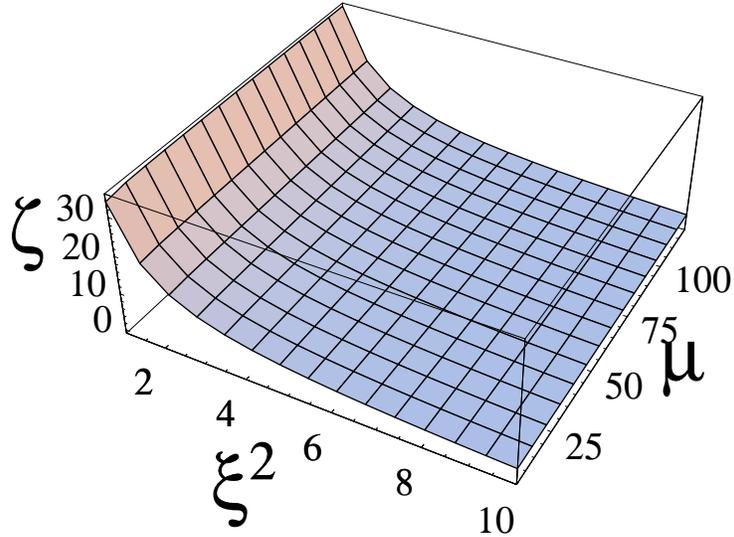,width=0.6\textwidth}}}
\caption{The self-interaction coupling $\zeta$ as a function 
of $\mu$ and $\xi$ for the fixed-point values
$g_\star=0.31$ and $\lambda_\star = 0.35$ evaluated from the RG 
equation for a scalar field. 
\label{fig2}}
\end{figure}

If instead ${\cal K}=\pm 1$ the only solution is for 
$\alpha =1, \beta=1,$ 
and we get the consistency conditions
\be
\mu = - \frac{9 \pi \xi^{-2} g_\star}{\zeta} +\frac{1}{4}
(6+\xi^2 \lambda_\star),  \;\;\;\;\;\;
\frac{\cal K}{A^2} = \frac{6 \pi \xi^{-2} g_\star}{\zeta}
+ {\xi^{2}\lambda_{\star}\over 2}, 
\;\;\;\;\; \phi_0 = \pm \sqrt {\frac{6}{\zeta}}. 
\ee 
For all cases, ${\cal K}=0$ and ${\cal K}\pm1$, 
$\Omega_{G}$ and $\Omega_\Lambda$ take constant values.  

\section{Concluding remarks}

The Lagrangian formulation of theories of gravity with variable 
Newton parameter has been considered both in classical \cite{[33]} 
and in quantum theory \cite{[34]} by a number of authors,
while the effective average action and renormalizability of
non-perturbative quantum gravity had been studied also
in Refs. \cite{[35],[36]}.

First, within the framework of relativistic theories of gravity 
with variable $G$ and $\Lambda$, 
we have arrived at the ADM Lagrangian (\ref{lag}), 
with coupling to a massless self-interacting scalar field as 
in Sec. III. Second, we have shown that, although the RG evolution
is not derivable in general from a Hamiltonian dynamics, 
the RG flow {\it at the fixed point} is consistent 
with a Lagrangian description of the basic running 
quantities $G$ and $\Lambda$ in a RW universe provided that we modify 
the standard Lagrangian for gravity by adding new terms which preserve
the standard form of momenta conjugate to the three-metric and ensure
a well-posed Hamilton variational problem. We only exploit the scaling of
scale factor, scalar field, $G$ and $\Lambda$ at the fixed point, which
is universal (in the sense of statistical mechanics), and find that
the desired fixed-point scaling actually exists even when the 
Lagrangian is allowed
to rule the behaviour of the Newton parameter.

Third, we have presented explicit solutions for pure gravity and a 
self-interacting scalar field coupled
to gravity in a RW Universe (see comments in Sec. III), finding
a class of power-law inflationary models which solve the horizon
problem of the cosmological standard model and can be used to test
against observation a fixed-point cosmology inspired by the
renormalization group but derived from a Lagrangian.

A challenging open problem is now the development of cosmological
perturbation theory starting from Lagrangians as in Eq. (10). This
could tell us whether formation of structure in the early universe
can also be accommodated within the framework of a Wilson-type
\cite{[37]} formulation of quantum gravity.

\appendix
\section{Hamiltonian analysis}

Our generalized framework with Lagrangian (10) engenders the familiar
primary constraints of general relativity, i.e.
\begin{equation}
\pi_{N} \equiv {\delta L \over \delta N_{,0}} \approx 0,
\label{(A1)}
\end{equation}
\begin{equation}
\pi^{i} \equiv {\delta L \over \delta {N_{i}}_{,0}} \approx 0,
\label{(A2)}
\end{equation}
with corresponding canonical Hamiltonian (for simplicity, we first
consider the pure gravity case)
\begin{equation}
H \equiv \int \Bigr(\pi_{N}N_{,0}+\pi^{i}{N_{i}}_{,0}
+\pi^{ij}h_{ij,0}+\pi_{G}G_{,0}\Bigr)d^{3}x-L,
\label{(A3)}
\end{equation}
where
\begin{equation}
\pi^{ij} \equiv {\delta L \over \delta h_{ij,0}}
=-{\sqrt{h}\over 16 \pi G}(K^{ij}-h^{ij}K),
\label{(A4)}
\end{equation}
\begin{equation}
\pi_{G} \equiv {\delta L \over \delta G_{,0}}
={\mu \sqrt{h}\over 8 \pi N G^{3}}\Bigr(G_{,0}-G_{,i}N^{i}\Bigr).
\label{(A5)}
\end{equation}
The resulting effective Hamiltonian (i.e. the canonical Hamiltonian 
plus a linear combination of primary constraints 
\cite{[28],[29]}) can be cast in the form
\begin{equation}
{\widetilde H} \equiv \int \Bigr(N {\cal H}_{0}+N_{i}{\cal H}^{i}
+\nu_{N}\pi_{N}+\nu_{i}\pi^{i}\Bigr)d^{3}x,
\label{(A6)}
\end{equation}
where $\nu_{N}$ and $\nu_{i}$ are Lagrange multipliers for primary
constraints, while ${\cal H}_{0}$ and ${\cal H}^{i}$ are the secondary
constraints obtained by preserving $\pi_{N}$ and $\pi^{i}$,
respectively. On defining the DeWitt (super-)metric on the space of
Riemannian metrics on $\Sigma$ \cite{[26]}:
\begin{equation}
G_{ijkl} \equiv {1\over 2\sqrt{h}}(h_{ik}h_{jl}+h_{il}h_{jk}
-h_{ij}h_{kl}),
\label{(A7)}
\end{equation}
one has
\begin{equation}
{\cal H}_{0} \equiv (16 \pi G)G_{ijkl}\pi^{ij}\pi^{kl}
-{\sqrt{h}\over 16 \pi G}{ }^{(3)}R+{\sqrt{h}\over 8\pi}
{\Lambda(G)\over G} 
+{4\pi G^{3}\over \mu \sqrt{h}}\pi_{G}^{2}
+{\mu \sqrt{h}\over 16 \pi G^{3}}h^{ij}G_{,i}G_{,j},
\label{(A8)}
\end{equation}
\begin{equation}
{\cal H}^{i} \equiv -2 \pi_{\; \; \; \mid j}^{ij}+h^{ij}G_{,j}\pi_{G}.
\label{(A9)}
\end{equation}

When gravity is coupled to an external scalar field ruled by the
Lagrangian (12), the secondary constraints read as
\begin{equation}
{\widetilde {\cal H}}_{0} \equiv {\cal H}_{0}
+{\sqrt{h}\over 2}\left[{\pi_{\phi}^{2}\over h}
+h^{ij}\phi_{,i}\phi_{,j}+2V(\phi)\right],
\label{(A10)}
\end{equation}
\begin{equation}
{\widetilde {\cal H}}^{i} \equiv {\cal H}^{i}+h^{ij}\phi_{,j}\pi_{\phi}.
\label{(A11)}
\end{equation}

In the cosmological models of Sec. III, the Lagrangian in Eq. (13) 
gives rise to the Hamiltonian (with $\pi_{a}$ the momentum conjugate
to the scale factor)
\begin{equation}
H= -{2\over 3}{\pi G \over a}\pi_{a}^{2}+{4\pi G^{3}\over \mu a^{3}}
\pi_{G}^{2}+{\pi_{\phi}^{2}\over 2a^{3}}-{3{\cal K}a\over 8\pi G}
+{\Lambda a^{3}\over 8\pi G}+a^{3}V(\phi).
\label{(A12)}
\end{equation}
The resulting Hamilton equations of motion are
\begin{equation}
{\dot a}=\left \{ a,H \right \}=-{4\pi G \over 3a}\pi_{a},
\label{(A13)}
\end{equation}
\begin{equation}
{\dot G}= \left \{ G,H \right \}={8\pi G^{3}\over \mu a^{3}}
\pi_{G},
\label{(A14)}
\end{equation}
\begin{equation}
{\dot \phi}=\left \{ \phi,H \right \}
={\pi_{\phi}\over a^{3}},
\label{(A15)}
\end{equation}
\begin{equation}
{\dot \pi}_{a}= \left \{ \pi_{a},H \right \}
=-{2\over 3}{\pi G \over a^{2}}\pi_{a}^{2}
+{12 \pi G^{3}\over \mu a^{4}}\pi_{G}^{2}
+{3 \pi_{\phi}^{2}\over 2a^{4}} 
+ {3{\cal K}\over 8 \pi G}
-{3 \Lambda a^{2}\over 8 \pi G}-3a^{2}V(\phi),
\label{(A16)}
\end{equation}
\begin{equation}
{\dot \pi}_{G}= \left \{ \pi_{G},H \right \}
={2\over 3}{\pi \over a}\pi_{a}^{2}
-{12 \pi G^{2}\over \mu a^{3}}\pi_{G}^{2}
-{3 {\cal K}a \over 8 \pi G^{2}} 
+ {\Lambda a^{3}\over 8 \pi G^{2}}
-{a^{3}\over 8 \pi G}{d\Lambda \over dG},
\label{(A17)}
\end{equation}
\begin{equation}
{\dot \pi}_{\phi}=\left \{ \pi_{\phi},H \right \}
=-a^{3}{dV\over d\phi}.
\label{(A18)}
\end{equation}
Equations (A13)--(A18) can be solved for given initial conditions
$a(0),G(0),\phi(0),\pi_{a}(0), \pi_{G}(0), \pi_{\phi}(0)$,
provided that such an initial data set satisfies the Hamiltonian
constraint $H \approx 0$ (cf. Eq. (17)).

Full agreement with the Euler--Lagrange equations (14)--(16) is
proved on taking the time derivative of Eqs. (A13)--(A15) and
then re-expressing the momenta $\pi_{a},\pi_{G},\pi_{\phi}$ 
and their first derivatives from Eqs. (A13)--(A18). For example,
Eq. (A13) implies that
$$
{\ddot a}=-{4\pi \over 3}\left({{\dot G}\over a}
-G{{\dot a}\over a^{2}}\right)\pi_{a}-{4\pi \over 3}
{G\over a}{\dot \pi}_{a},
$$
and the insertion of Eqs. (A13) and (A16) yields eventually Eq. (14)
of Sec. III.

\acknowledgments
The authors are grateful
to the INFN for financial support. 
The work of G. Esposito has been partially supported by
PRIN 2002 ``Sintesi.'' Comments and criticism of 
Hans J. Matschull and Martin Reuter have been very helpful 
in the course of completing our research. Correspondence with
Sergei Odintsov is also gratefully acknowledged.

\end{document}